\documentclass[aps,pra,superscriptaddress,showpacs,floatfix,lengthcheck,nofootinbib,longbibliography,twocolumn]{revtex4-1}   
\usepackage[T1]{fontenc}
\usepackage[utf8]{inputenc}
\usepackage{amsmath}
\usepackage{amssymb}
\usepackage{graphicx}
\usepackage{color}
\usepackage[pdfusetitle,colorlinks,allcolors=blue]{hyperref}
\usepackage{microtype}
\usepackage{braket}

\def\equationautorefname~#1\null{Eq.~(#1)\null}

\begin{document}
\title{Permutational symmetry for identical multi-level systems: a second
quantized approach}

\author{Rui E. F. Silva}
\email{ruiefdasilva@gmail.com}
\affiliation{Instituto de Ciencia de Materiales de Madrid, Consejo Superior
de Investigaciones Científicas (ICMM-CSIC), Madrid, Spain}
\affiliation{Departamento de Física Teórica de la Materia Condensada and
Condensed Matter Physics Center (IFIMAC), Universidad Autónoma de
Madrid, E-28049 Madrid, Spain}

\author{Johannes Feist}
\email{johannes.feist@uam.es}
\affiliation{Departamento de Física Teórica de la Materia Condensada and
Condensed Matter Physics Center (IFIMAC), Universidad Autónoma de
Madrid, E-28049 Madrid, Spain}

\begin{abstract}
We develop a framework that provides a straightforward approach to fully
exploit the permutational symmetry of identical multi-level systems. By taking
into account the permutational symmetry, we outline a simple scheme that allows
to map the dynamics of $N$ identical $d$-level systems to the dynamics of $d$
bosonic modes with $N$ particles, achieving an exponential reduction on the
dimensionality of the problem in a simple and straightforward way. In
particular, we consider the Lindblad dynamics of several identical multi-level
systems interacting with a common subsystem under the action of collective
dissipation terms. 
\end{abstract}
\maketitle

\section{Introduction}

When dealing with a collection of $N$ $d$-level systems, a well-known problem is
the so-called curse of dimensionality, i.e., the fact that the dimension of the
Hilbert space scales exponentially as $d^{N}$. However, in many different
physical phenomena, such as lasing \citep{sculybook,richter2015numerically},
phase transitions \citep{wang1973phase,walls1978non,gegg2018superradiant},
superradiance \citep{dicke1954coherence,garraway2011dicke}, strong coupling with
organic molecules \citep{herrera2016cavity,feist2018polaritonic} and microwave
photonics \citep{putz2017spectral}, the theoretical modeling usually assumes
that the emitters are identical. In these situations, permutational symmetry of
the $N$ $d$-level systems can be used to greatly reduce the complexity of the
problem. This was addressed in the works of Gegg \emph{et
al.}~\citep{gegg2016efficient, gegg2017psiquasp, gegg2017identical}, Shammah
\emph{et al.}~\citep{piqs_paper} and Kirton \emph{et
al.}~\citep{kirton2017suppressing, kirton2018superradiant}. In these works, the
dynamics of an open quantum system composed of several identical emitters
interacting with a common subsystem under the action of individual, but
identical, collapse operators is considered. By exploiting the permutational
symmetry of the density matrix in the symmetrized Liouville space, a huge
reduction in the complexity of the problem is achieved, allowing calculations
for larger numbers of emitters than possible otherwise. These efforts were
conducted for an ensemble of multi-level systems~\citep{gegg2016efficient,
gegg2017psiquasp, gegg2017identical} and specialized for the case of two-level
systems~\citep{piqs_paper, kirton2017suppressing, kirton2018superradiant}. In
the absence of individual dephasing operators and for appropriate initial
states, one can further restrict the Hilbert space to the totally symmetric
subspace~\citep{gegg2016efficient,gegg2017identical}. In the case of $2$-level
systems, the construction of the totally symmetric subspace can be achieved by
using the Dicke basis, restricting the Hilbert space to the highest super-spin
subspace~\citep{dicke1954coherence, garraway2011dicke}.

In this work, we notice that working in the totally symmetric subspace is
completely equivalent to restricting the possible states to bosonic many-body
states. Therefore, by applying the rules of second quantization for bosons, we
achieve the reduction to the totally symmetric subspace in a simple and
straightforward way, mapping the dynamics of $N$ identical $d$-level systems to
the dynamics of $d$ bosonic modes with $N$ particles. 

\section{Theory}

\begin{figure*}
\begin{centering}
\includegraphics[width=2\columnwidth]{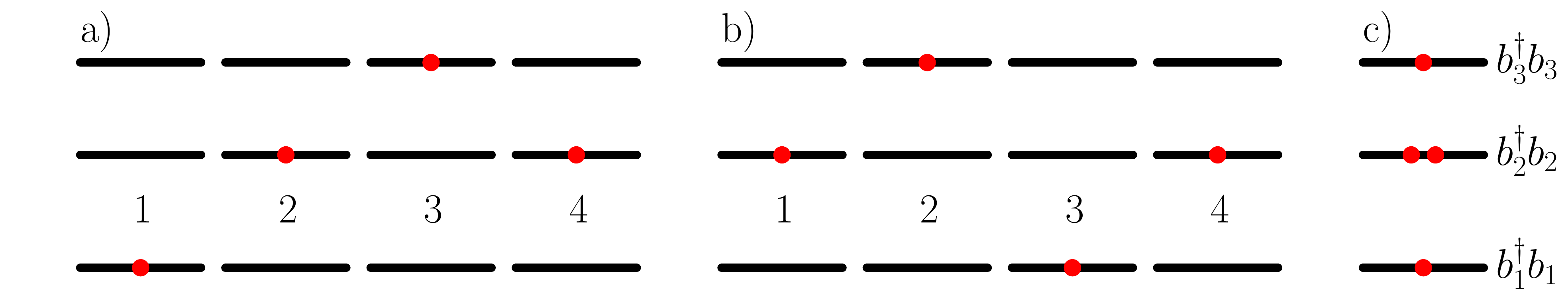}
\par\end{centering}
\caption{\label{fig:1}(a,b) Schematic representation of two product states
of four 3-level systems. (a) $\left|1\right\rangle _{1}\left|2\right\rangle _{2}\left|3\right\rangle _{3}\left|2\right\rangle _{4}$
and (b) $\left|2\right\rangle _{1}\left|3\right\rangle _{2}\left|1\right\rangle _{3}\left|2\right\rangle _{4}$.
If we restrict   dynamics to the totally symmetric subspace, these
two states will be always in a symmetric superposition that can be
correctly taken into account by using many-body bosonic states, (c)
$\frac{1}{\sqrt{2!}}b_{3}^{\dagger}b_{2}^{\dagger}b_{2}^{\dagger}b_{1}^{\dagger}\left|\mathrm{vac}\right\rangle $. }
\end{figure*}

We start by considering the dynamics of $N$ identical $d$-level systems
interacting with a common subsystem and under the action of collective
dissipation terms, described by the Lindblad master equation,
\begin{equation}
\dot{\rho}=-i\left[H,\rho\right]+\sum_{i}\mathcal{L}_{C_{i}}\left[\rho\right],
\end{equation}
where $H$ is the Hamiltonian, possibly time-dependent, and $\mathcal{L}_{C_{i}}\left[\rho\right]=C_{i}\rho C_{i}^{\dagger}-\frac{1}{2}\left(C_{i}^{\dagger}C_{i}\rho+\rho C_{i}^{\dagger}C_{i}\right)$
is the Lindblad dissipator for the collapse operator $C_{i}$. In
the following, we assume the Hamiltonian, $H$, to be invariant under
any permutation of the $d$-level systems. We also restrict the collapse
operators, $C_{i}$, to collective operators that are also invariant
under any permutation of the $d$-level systems. For the case where
the collapse operators may act locally on each $d$-level system,
we must construct and work on the symmetrized Liouville space and
this was taken into account in \citep{gegg2016efficient,gegg2017psiquasp,gegg2017identical,piqs_paper,kirton2017suppressing,kirton2018superradiant}.
In this work, we restrict to the case where both the Hamiltonian and
the collapse operators are invariant under any permutation of the
emitters. 

We may define the symmetrization operator, 
\begin{equation}
S=\frac{1}{N!}\sum_{\pi}P_{\pi}
\end{equation}
where $P_{\pi}$ is a permutation operator and $\pi$ runs over all possible
permutations of the $d$-level systems. A permutationally invariant operator,
$O$, then satisfies $[S,O]=0$. We also assume that the initial state,
$\rho_{0}=\sum_{i}p_{i}\left|\psi_{i}\right\rangle \left\langle
\psi_{i}\right|$, is a totally symmetric state, i.e.,
$S\left|\psi_{i}\right\rangle =\left|\psi_{i}\right\rangle $ for all
$\left|\psi_{i}\right\rangle $. Since both the Hamiltonian, $H$, and the
collapse operators, $C_{i}$, are permutationally invariant operators, when
solving the Lindblad dynamics, the density matrix will always remain in the
totally symmetric subspace. This may be used to substantially reduce the
dimensionality of the problem. As already noticed in \citep{gegg2016efficient,
gegg2017identical}, this reduction of dimensionality is even larger than the one
obtained by working in the symmetrized Liouville space. We must stress that the
difference in the restrictions imposed in this work and the symmetrized
Liouville space approach is that here, each collapse operator $C_{i}$ must be
permutationally invariant, whereas in Refs.~\citep{gegg2016efficient,
gegg2017psiquasp, gegg2017identical, piqs_paper, kirton2017suppressing,
kirton2018superradiant} only the sum of all collapse operators, $\sum_i C_{i}$,
must be permutationally invariant.

We can write any permutationally invariant $M$-body operator acting solely on
the emitters as
\begin{equation}
O_{\mathrm{em}}^{M} = \frac{1}{M!} \sideset{}{'}\sum_{i_{1}\ldots i_{M}}^{N}\sum_{\substack{\alpha_{1}\ldots\alpha_{M}\\
\beta_{1}\ldots\beta_{M}}}^{d} V_{\alpha_{1}\ldots\alpha_{M}}^{\beta_{1}\ldots\beta_{M}} \sigma_{\beta_1,\alpha_1}^{i_1}\ldots\sigma_{\beta_M,\alpha_M}^{i_M} \label{eq:emitter_op_symmetric-1}
\end{equation}
where $\sigma_{\beta,\alpha}^{j}=\left|\beta\right\rangle _{j}\left\langle
\alpha\right|_{j}$ is an operator acting on emitter $j$, and the primed sum
indicates that all indices $i_1,\ldots,i_M$ have to be distinct. Using this
definition, any permutationally invariant operator acting on the emitters and
containing up to $M$-body terms can be written as
\begin{equation}
O_{\mathrm{em}} = \sum_{J=1}^{M}O_{\mathrm{em}}^{J}.\label{eq:emitter_op_symmetric}
\end{equation}
In general, a permutationally invariant operator that may act on the emitters
and on a common subsystem can be written as the sum of three terms
\begin{equation}
O = O_{\mathrm{em}}+O_{\mathrm{em-sub}}+O_{\mathrm{sub}},\label{eq:general_operator_symmetry_first_quant}
\end{equation}
where $O_{\mathrm{em}}$ ($O_{\mathrm{sub}}$) is an operator acting only on the
emitters (subsystem). The interaction term, $O_{\mathrm{em-sub}}$, can be
written as $O_{\mathrm{em-sub}}=\sum_{q}A_{q}B_{q}$, where $A_{q}$ acts solely
on the emitters and must have the form of \autoref{eq:emitter_op_symmetric}
and $B_{q}$ is an operator acting on the common subsystem. 

At this point, one may realize that if the state of the system is restricted the
totally symmetric Hilbert space, one can take advantage of all the formalism of
second quantization for bosons, for which the many-body states are automatically
restricted to the totally symmetric Hilbert space. This can be done by applying
the rules of second quantization~\citep{flensberg_book} and mapping all the
operators to a second quantized form. For instance,
\autoref{eq:emitter_op_symmetric-1} becomes
\begin{equation}
    O_{\mathrm{em}}^{M} = \frac{1}{M!} \sum_{\substack{\alpha_{1}\ldots\alpha_{M}\\
     \beta_{1}\ldots\beta_{M}}}^{d} V_{\alpha_{1}\ldots\alpha_{M}}^{\beta_{1}\ldots\beta_{M}} b_{\beta_{1}}^{\dagger}\ldots b_{\beta_{M}}^{\dagger}b_{\alpha_{1}}\ldots b_{\alpha_{M}}
\end{equation}
where $b_{\alpha}^{\dagger}$ and $b_{\alpha}$ are the bosonic creation and
annihilation operators of an emitter in state $\alpha$. The recipe then simply
consists in applying these rules to all the relevant operators, i.e., the
Hamiltonian, the collapse operators and all the desired observables, giving a
Hamiltonian that can be easily implemented with standard quantum optics packages
such as QuTiP~\cite{johansson2012nation}. For the typical cases where the number
of emitters $N$ is fixed, one can restrict the Hilbert space to the $N$-particle
subspace, for which $\left\langle \sum_{\alpha=1}^{d}
b_{\alpha}^{\dagger}b_{\alpha}\right\rangle = N$. Similarly, for the initial
state, $\rho_{0}=\sum_{i}p_{i}\left|\psi_{i}\right\rangle \left\langle
\psi_{i}\right|$, each $\left|\psi_{i}\right\rangle$ has to be mapped to its
second quantized version, i.e., expressed in the Fock space.

This approach thus solves the dynamics of $N$ permutationally invariant
$d$-level systems by treating it as the dynamics of $N$ bosons in a system with
$d$ modes. It has the usual advantages of a second quantized formulation. In
particular, it is not necessary to explicitly construct a totally symmetric
subspace, and the correct symmetry enhancement factors are automatically encoded
within and obtained from the bosonic operator algebra.

\section{Results}

The study of few-level emitters interacting with light is at the core of our
understanding of light-matter interaction. In principle, the problem of
light-matter interaction can be fully understood within the laws of quantum
electrodynamics (QED)~\citep{cohen_qed_book}.
However, for practical applications in fields such as cavity QED, quantum
optics, quantum nanophotonics, and quantum plasmonics, a very common assumption
is that matter degrees of freedom can be described using only a few levels and
that the interaction with light is dominated by a single mode of the
electromagnetic field. In the case of two-level systems, this leads to the well
known Rabi~\citep{rabi1937space}, Dicke~\citep{dicke1954coherence},
Jaynes-Cummings~\citep{jaynes1963comparison} and
Tavis-Cummings~\citep{tavis1968exact} models, which differ in the number of
emitters and the use of the rotating wave approximation. Nevertheless, even when
dealing with an ensemble of few-level systems, the exponential scaling of the
Hilbert space dimension quickly makes the problem intractable and symmetry
considerations must be taken into account to reduce the dimension of the
problem. In the context of the interaction of an ensemble of identical two-level
systems with a cavity mode, superradiance can be observed. To study
superradiance, it is useful to rewrite the Hamiltonian using spin operators and
use the so-called Dicke basis~\citep{garraway2011dicke}. In the following, we
demonstrate our approach for several examples within this context.

\subsection{Tavis-Cummings model}

As a first example to illustrate our approach, we will apply it to the
Tavis-Cummings model~\citep{tavis1968exact}. For this relatively simple example,
we show explicitly that the state space and matrix elements within the second
quantized picture are the same as in conventional approaches. The Tavis-Cummings
Hamiltonian is given by
\begin{align}
H_{\mathrm{TC}} & =\sum_{j=1}^{N}\frac{\omega_{0}}{2}\left(\sigma_{\mathrm{e},\mathrm{e}}^{j}-\sigma_{\mathrm{g},\mathrm{g}}^{j}\right)+\omega_{\mathrm{c}}a_{\mathrm{c}}^{\dagger}a_{\mathrm{c}}\nonumber \\
 & +g\sum_{j=1}^{N}\left(a_{\mathrm{c}}\sigma_{\mathrm{e},\mathrm{g}}^{j}+a_{\mathrm{c}}^{\dagger}\sigma_{\mathrm{g},\mathrm{e}}^{j}\right)\label{eq:TC_first}
\end{align}
where $\mathrm{g}$ ($\mathrm{e}$) stands for the ground (excited) state, $N$ is
the number of two-level systems, $\omega_{0}$ ($\omega_{\mathrm{c}}$) is the
two-level system (cavity) energy and $a_{c}$ is the bosonic annihilation
operator for the cavity. A standard approach is to rewrite the above Hamiltonian
using spin operators and work in the Dicke basis \citep{garraway2011dicke},
where $S_{+}=\sum_{j=1}^{N}\sigma_{\mathrm{e},\mathrm{g}}^{j}$, $S_{-} =
\left(S_{+}\right)^{\dagger}$ and $S_{z} = \sum_{j=1}^{N} \frac{1}{2}
\left(\sigma_{\mathrm{e},\mathrm{e}}^{j}-\sigma_{\mathrm{g},\mathrm{g}}^{j}\right)$.
The Hamiltonian can then be written as
\begin{equation}
H_{\mathrm{TC}}=\omega_{0} S_{z} + \omega_{\mathrm{c}} a_{\mathrm{c}}^{\dagger} a_{\mathrm{c}} + 
g \left(a_{\mathrm{c}} S_{+} + a_{\mathrm{c}}^{\dagger} S_{-}\right),
\end{equation}
and the emitter states are $\left|s,m\right\rangle$, where $s$ and $m$ are the
quantum numbers associated to $S^{2}$ and $S_{z}$. The totally symmetric
subspace is then the highest spin subspace, where $s=N/2$.

If we instead apply our approach and second quantize \autoref{eq:TC_first}, we
obtain
\begin{equation}
H_{\mathrm{TC}} = \frac{\omega_{0}}{2}\left(b_{\mathrm{e}}^{\dagger}b_{\mathrm{e}}-b_{\mathrm{g}}^{\dagger}b_{\mathrm{g}}\right)+\omega_{\mathrm{c}}a_{\mathrm{c}}^{\dagger}a_{\mathrm{c}}+g\left(a_{\mathrm{c}}b_{\mathrm{e}}^{\dagger}b_{\mathrm{g}}+a_{\mathrm{c}}^{\dagger}b_{\mathrm{g}}^{\dagger}b_{\mathrm{e}}\right).\label{eq:TC_second}
\end{equation}
In order to see that both approaches are completely equivalent in the totally
symmetric subspace ($s=N/2$), we examine the matrix elements of $S_{-}$. The action of
$S_{-}$ on a Dicke state is
\begin{equation} S_{-}\left|s,m\right\rangle =
\sqrt{s(s+1)-m(m-1)}\left|s,m-1\right\rangle. \label{eq:s_minus_formula}
\end{equation}
Here, $m=N_{\mathrm{exc}}-N/2$ is directly related to the number of excited
emitters, $N_{\mathrm{exc}}$. 

For the second quantized version, $S_{-}$ maps to $b_{g}^{\dagger}b_{e}$,
which acts on the Fock states $\left|n_{g},n_{e}\right\rangle $,
where $n_{g}=N-N_{\mathrm{exc}}$ and $n_{e}=N_{\mathrm{exc}}$, as
\begin{equation}
b_{g}^{\dagger}b_{e}\left|n_{g},n_{e}\right\rangle =\sqrt{(n_{g}+1)n_{e}}\left|n_{g}+1,n_{e}-1\right\rangle .\label{eq:bgbe_TC}
\end{equation}
The Dicke state $\left|s=N/2,m=N_{\mathrm{exc}}-N/2\right\rangle$ is equal to
the Fock state $\left|N-N_{\mathrm{exc}},N_{\mathrm{exc}}\right\rangle$, and
comparing \autoref{eq:s_minus_formula} and \autoref{eq:bgbe_TC} shows that the
matrix elements are indeed equal. The equality can also easily be checked for
$S_{z} \equiv \frac{1}{2} \left(b_{\mathrm{e}}^{\dagger}b_{\mathrm{e}} -
b_{\mathrm{g}}^{\dagger}b_{\mathrm{g}}\right)$. Therefore, within the totally
symmetric subspace, it is completely equivalent to work with either of the two
Hamiltonians, \autoref{eq:TC_first} or \autoref{eq:TC_second}.

\subsection{Holstein-Tavis-Cummings model}

The field of molecular polaritonics and polaritonic chemistry
\citep{Hertzog2019, Herrera2020, climent2021cavity, Garcia-Vidal2021,
Fregoni2021Perspective, Sanchez-Barquilla2022Perspective} studies how to
manipulate and use the changes in electronic and vibrational structure and
dynamics of molecules under strong coupling with confined modes of light. Since
molecules are complex systems with significant internal structure due to
rovibrational (nuclear) motion, describing them as two-level systems is often
not a good approximation. At the same time, the influence of individual collapse
operators acting on each molecule can often be neglected. On the one hand, their
individual radiative decay (on scales of nanoseconds) is often much slower than
the dynamics of interest. On the other hand, the influence of the vibrational
modes that is sometimes included through a pure-dephasing Lindblad term (which
has to be replaced by a more careful treatment under strong light-matter
coupling to prevent unphysical effects~\cite{del2015quantum}) can be much better
described by treating some vibrational modes (or superpositions of them
corresponding to so-called reaction coordinates) explicitly, which allows
neglecting the other ones at reasonably short
timescales~\cite{silva2020polaritonic, zhao2020impact}. These considerations
apply especially for organic molecules interacting with a plasmonic
nanocavity~\citep{chikkaraddy2016single, ojambati2019quantum}, since their
ultrafast loss is typically the dominant decay channel in the system. Explicit
inclusion of nuclear degrees of freedom also allows to represent many effects
that cannot be understood within a two-level system
description~\citep{herrera2016cavity,galego2015cavity}. A workhorse in this
field is the so-called Holstein-Tavis-Cummings model~\citep{herrera2016cavity},
in which the molecule is approximated using the Holstein model, i.e., two
displaced harmonic oscillators for the electronic ground and excited states.
Therefore, when dealing with molecular polaritonics, it is common to face
situations where one needs to solve the dynamics of identical multi-level
systems without any individual collapse operator.

The Holstein-Tavis-Cummings Hamiltonian can be written as
\begin{gather}
H_{\mathrm{HTC}} = \omega_{\mathrm{c}} a_{\mathrm{c}}^{\dagger} a_{\mathrm{c}} + \sum_{i=1}^{N_{\mathrm{mol}}} H_{\mathrm{mol}}^{(i)} + 
\sum_{i=1}^{N_{\mathrm{mol}}} H_{\mathrm{cav-mol}}^{(i)},\label{eq:HTC_HAM}\\
H_{\mathrm{mol}}^{(i)} = \omega_{\mathrm{e}} \sigma_{i}^{+}\sigma_{i}^{-}+\omega_{\mathrm{v}}c_{i}^{\dagger}c_{i}-\lambda_{\mathrm{v}}\sigma_{i}^{+}\sigma_{i}^{-}\left(c_{i}^{\dagger}+c_{i}\right)\\
H_{\mathrm{cav-mol}}^{(i)} = g \left(\sigma_{i}^{+}a_{\mathrm{c}} + a_{\mathrm{c}}^{\dagger} \sigma_{i}^{-}\right),
\end{gather}
where $\sigma_{i}^{+}$ ($\sigma_{i}^{-}$) is the raising (lowering) operator for
the electronic state in molecule $i$ with excitation energy
$\omega_{\mathrm{e}}$, whereas $c_{i}$ is the annihilation operator for the
vibrational mode in molecule $i$, with frequency $\omega_{\mathrm{v}}$ and
exciton--phonon coupling strength $\lambda_{\mathrm{v}}$. The cavity is
described through the photon annihilation (creation) operators $a_{\mathrm{c}}$
($a_{\mathrm{c}}^{\dagger}$), with photon energy $\omega_{\mathrm{c}}$. In
addition to the coherent dynamics described by the Hamiltonian, the cavity mode
decays with rate $\gamma_{\mathrm{c}}$, described by a standard Lindblad decay
operator $C=\sqrt{\gamma_{\mathrm{c}}} a_{\mathrm{c}}$.

The Holstein-Tavis-Cummings Hamiltonian can be rewritten in terms of the
eigenstates of the single-molecule Hamiltonian, 
\begin{equation}
H_{\mathrm{mol}}^{(i)} = \sum_{s=\mathrm{g},\mathrm{e}} \sum_{\nu} \omega_{\mathrm{s},\nu} |\mathrm{s},\nu\rangle_{i} \langle \mathrm{s},\nu|_{i},
\end{equation}
which are labeled as $|\mathrm{g},\nu\rangle_{i}$ and
$|\mathrm{e},\nu\rangle_{i}$ for vibrational sublevel $\nu$ in the electronic
ground and excited state, respectively. Their corresponding energies are
$\omega_{\mathrm{g},\nu}=\omega_{\mathrm{v}} \nu$ and
$\omega_{\mathrm{e},\nu}=\omega_{\mathrm{e}} +
\omega_{\mathrm{v}}\nu-\lambda_{\mathrm{v}}^{2}/\omega_{\mathrm{v}}$. In this basis,
the light-matter interaction operator is given by
\begin{equation}
H_{\mathrm{cav-mol}}^{(i)} = g \sum_{\nu\nu'}\left(\text{\ensuremath{a_{\mathrm{c}}F_{\nu\nu'}\left|\mathrm{e},\nu\right\rangle _{i}\left\langle \mathrm{g},\nu'\right|_{i}}}+\mathrm{H.c.}\right),
\end{equation}
where $F_{\nu\nu'}=\left\langle
\mathrm{e},\nu\left|\sigma_{i}^{+}\right|\mathrm{g},\nu'\right\rangle $ is a
vibrational overlap integral or Franck-Condon factor and can be analytically
obtained.

\begin{figure}
    \includegraphics[width=\linewidth]{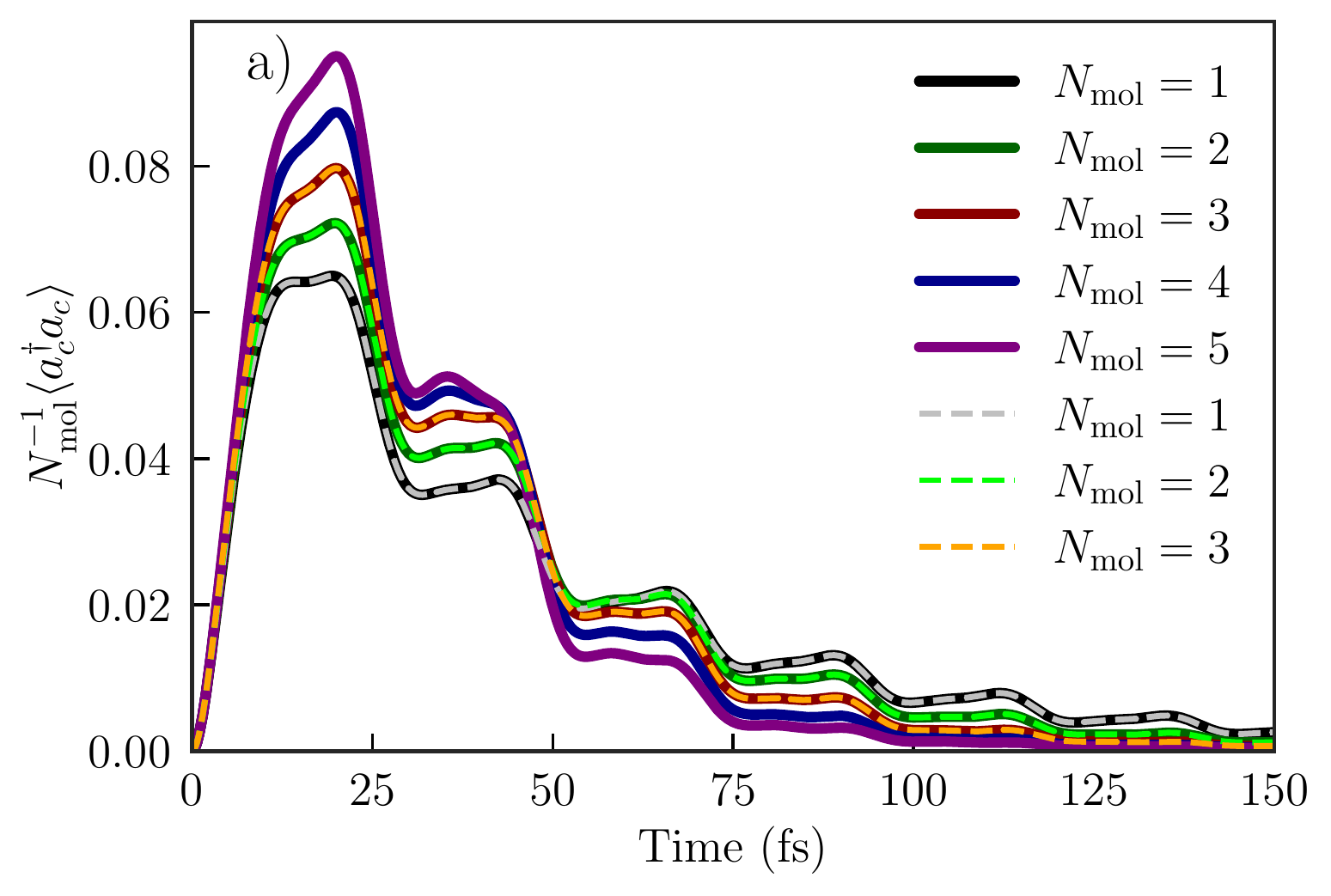}
    \includegraphics[width=\linewidth]{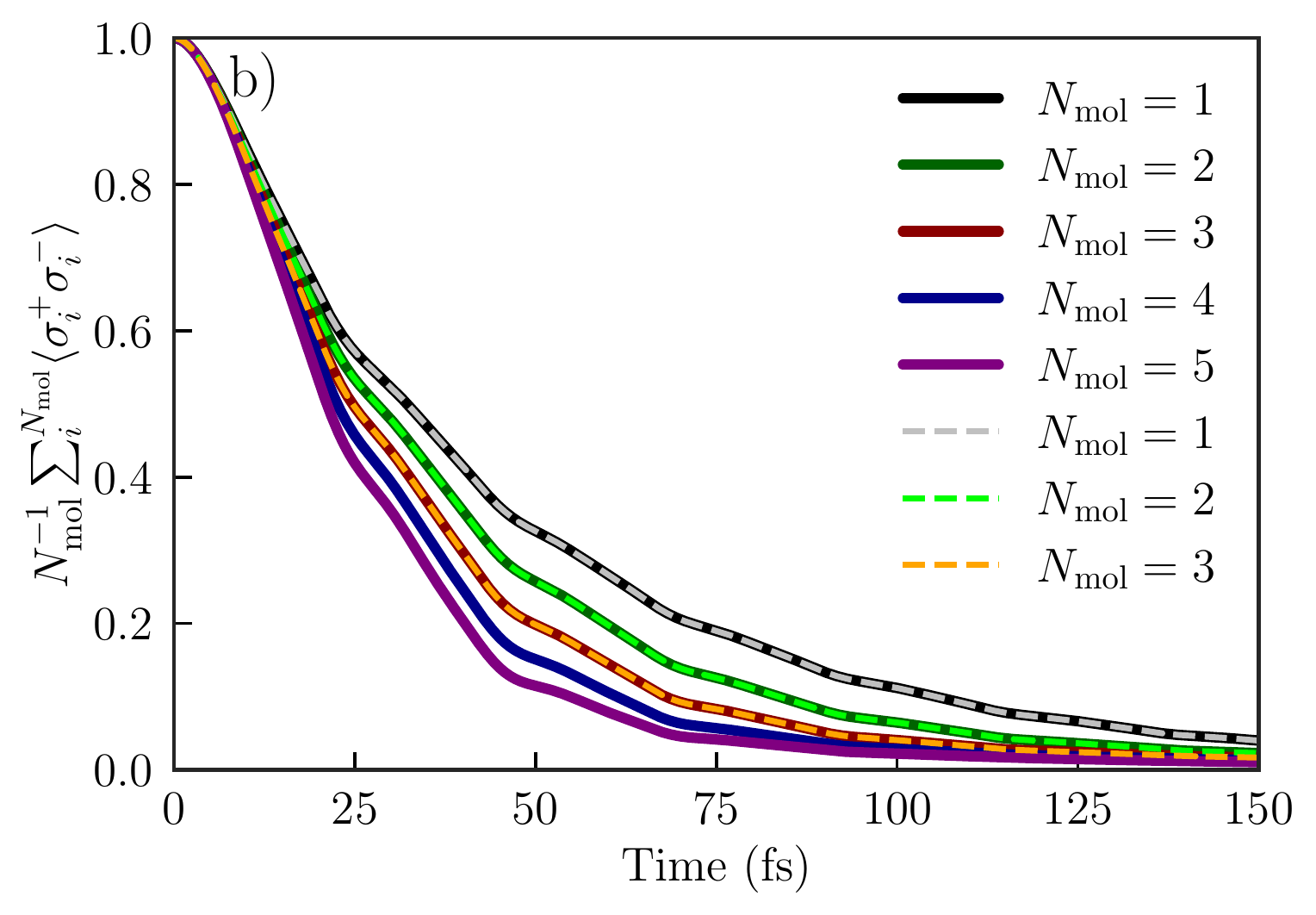}
    \caption{\label{fig:HTC}\label{fig:HTC_a}(a) Normalized cavity population
    $\left\langle a_{\mathrm{c}}^{\dagger} a_{\mathrm{c}} \right\rangle
    N_{\mathrm{mol}}^{-1}$ in the Holstein-Tavis-Cummings model, for
    $N_{\mathrm{mol}}=1,2,3,4,5$ calculated both within the second-quantization
    approach (full lines) and without exploiting permutational symmetry (dashed
    lines). See main text for parameters. \label{fig:HTC_b}(b) The same for the
    electronic excited state population,
    $\sum_{i}^{N_{\mathrm{mol}}}\left\langle
    \sigma_{i}^{+}\sigma_{i}^{-}\right\rangle N_{\mathrm{mol}}^{-1}$.}
\end{figure}

Since both the Hamiltonian and collapse operators are permutationally invariant, we can map \autoref{eq:HTC_HAM} to its
second quantized form as long as the initial state is fully symmetric. This gives
\begin{multline}
    H_{\mathrm{HTC}} = \omega_{\mathrm{c}} a_{\mathrm{c}}^{\dagger} a_{\mathrm{c}} 
    + \sum_{s=\mathrm{g},\mathrm{e}} \sum_{\nu} \omega_{\mathrm{s},\nu} b_{\mathrm{s},\nu}^{\dagger} b_{\mathrm{s},\nu}\\
    + g \sum_{\nu\nu'}\left(\text{\ensuremath{a_{\mathrm{c}}F_{\nu\nu'}b_{\mathrm{e},\nu}^{\dagger}b_{\mathrm{g},\nu'}}} + \mathrm{H.c.}\right).
\end{multline}

In the following, we choose parameter values typical for organic molecules such
as anthracene coupled to nanoplasmonic cavities~\cite{silva2020polaritonic},
with $\omega_{\mathrm{e}}=3.5\,$eV, $\omega_{\mathrm{v}}=0.182\,$eV,
$\lambda_{\mathrm{v}}=0.096\,$eV, $\gamma_{\mathrm{c}}=0.2\,$eV, $g=0.035\,$eV. We set the
cavity photon energy to be on resonance with the emission peak of the molecule,
$\omega_{\mathrm{c}} = \omega_{\mathrm{e}} - 2\lambda_{\mathrm{v}}^{2} /
\omega_{\mathrm{v}}$. The initial state is chosen to be the fully inverted
state, i.e., the state where all molecules are instantaneously excited to the
electronic excited state by a vertical Franck-Condon transition. 

In \autoref{fig:HTC}(a,b), we show the numerical results for the dynamics for
the cavity and excited state population, respectively. In the basis truncation
for the single-molecule Hilbert space, we include the 6 lowest vibrational
states for the ground state and the 4 lowest vibrational states for the
electronic excited state, which gives converged results. Within the
second quantization approach, we show results up to $N_{\mathrm{mol}}=5$, while
for the results without resorting to any permutational symmetry, we show results
up to $N_{\mathrm{mol}}=3$. 

The exciton population in \autoref{fig:HTC}(b) displays a clear
enhancement of the spontaneous emission due to Dicke
superradiance~\cite{dicke1954coherence} as the number of emitters is increased.
Furthermore, a modulation of the decay rate with a period of about $22\,$fs can
be observed. This modulation is more clearly visible in the cavity population,
see \autoref{fig:HTC}(a), and is a signature of the vibrational
motion~\cite{silva2020polaritonic}, which has a period of $T_\mathrm{v} = 2 \pi
/ \omega_{\mathrm{v}} = 22.7\,$fs.

As expected, both approaches are completely equivalent. However, while the
Hilbert space for the brute-force approach reaches size $N_\mathrm{Hilb} = 4000$
for three molecules, it only has size $N_\mathrm{Hilb} = 220$ within the second
quantization approach exploiting the permutational symmetry. For five molecules,
this advantage improves to $N_\mathrm{Hilb} = 12012$ versus $N_\mathrm{Hilb} =
600000$. Here, it should be noted that the size of the density matrix that is
propagated in the Lindblad master equation is $N_\mathrm{Hilb} \times
N_\mathrm{Hilb}$, while the Liouvillian superoperator describing this evolution
can be formally treated as a $N_\mathrm{Hilb}^2 \times N_\mathrm{Hilb}^2$
matrix.

\subsection{Three-level systems}

\begin{figure}
\includegraphics[width=\linewidth]{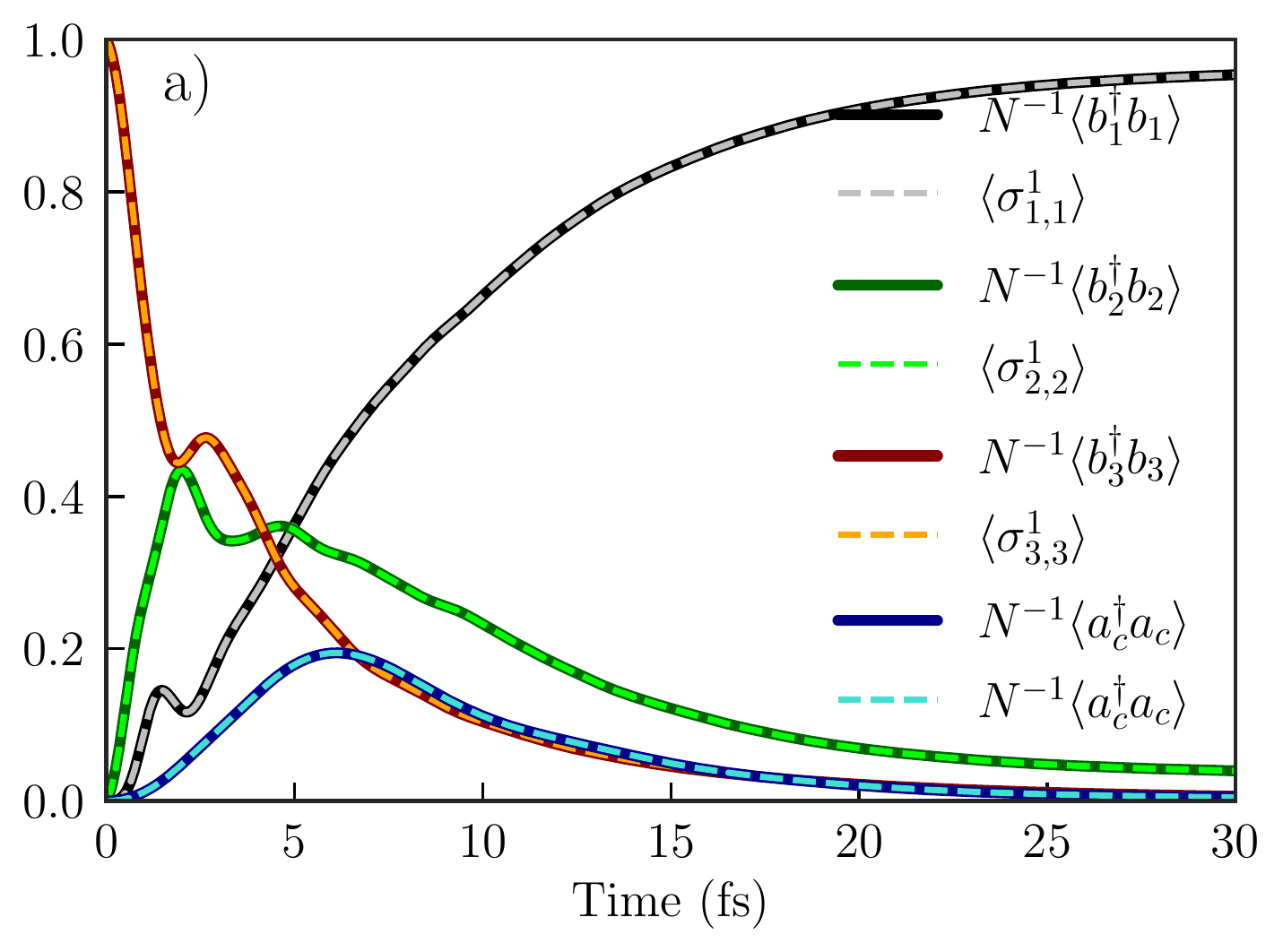}
\includegraphics[width=\linewidth]{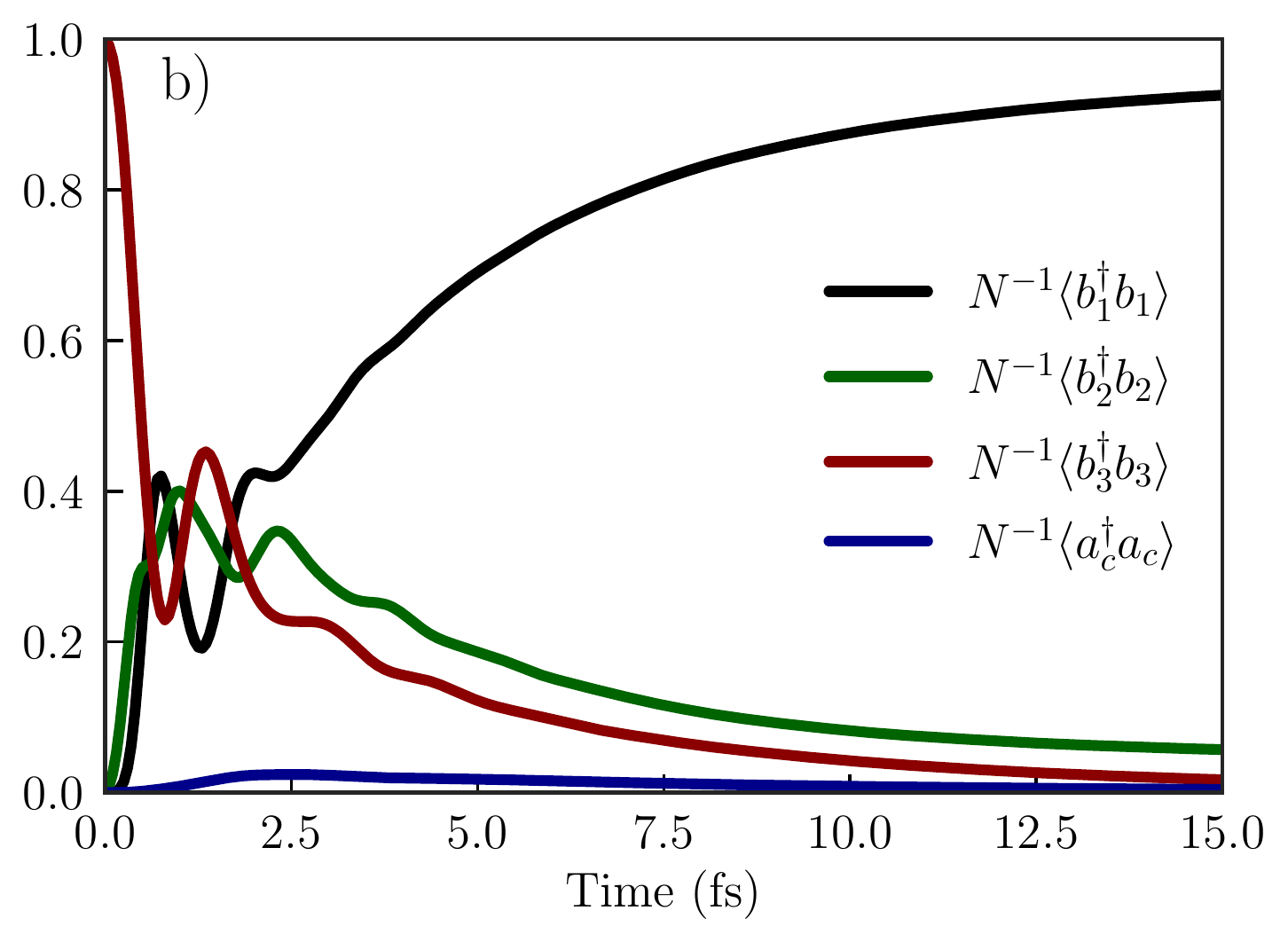}
\caption{\label{fig:2} (a) Dynamics for $N=5$ three-level emitters calculated
within the second quantized approach (full lines) and without using the
permutational symmetry (dashed lines). See main text for parameters. (b) The
same for $N=17$ emitters, which is only possible with reasonable effort when
using the second quantized approach.}
\end{figure}

\begin{figure}
\includegraphics[width=\linewidth]{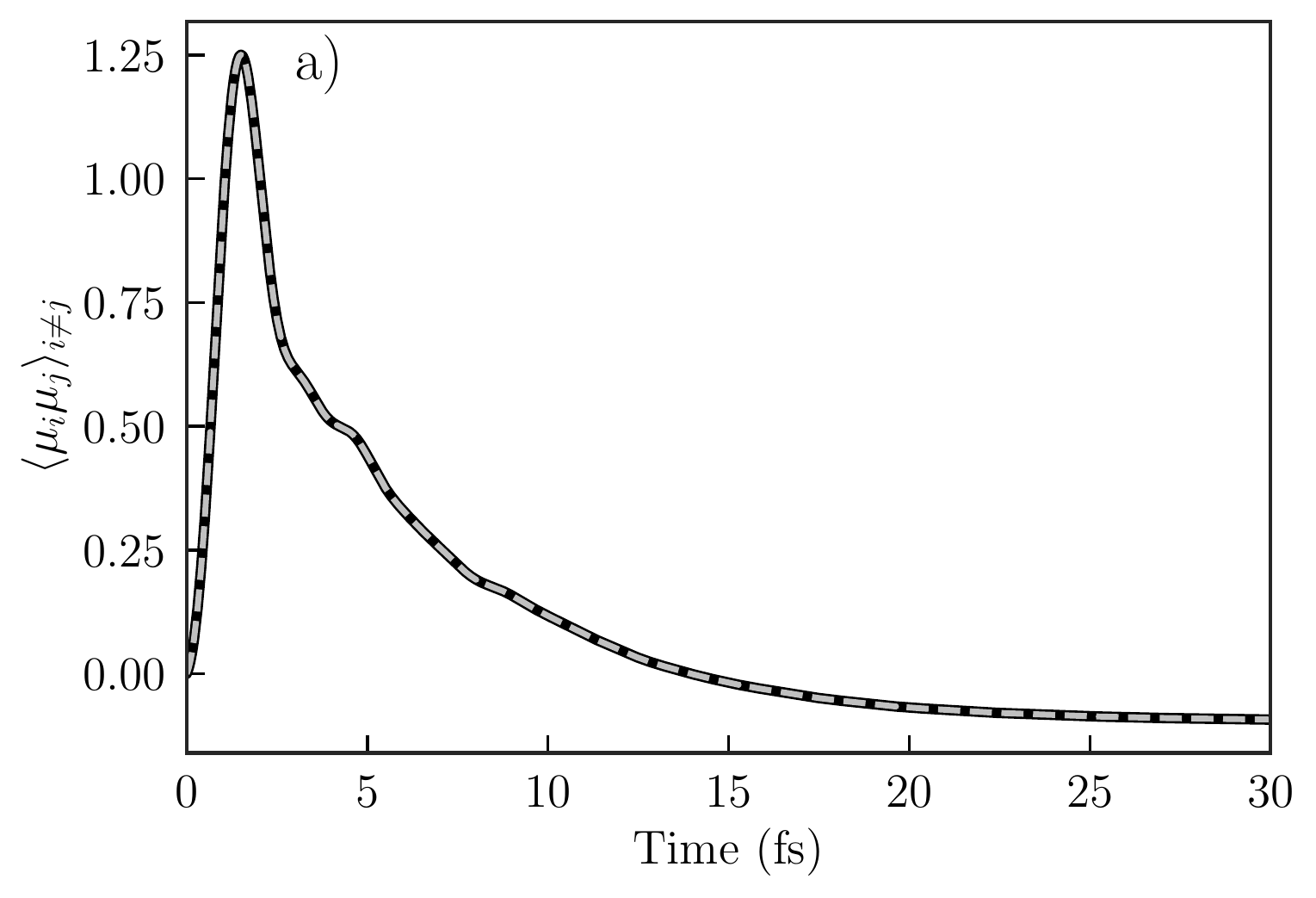}
\includegraphics[width=\linewidth]{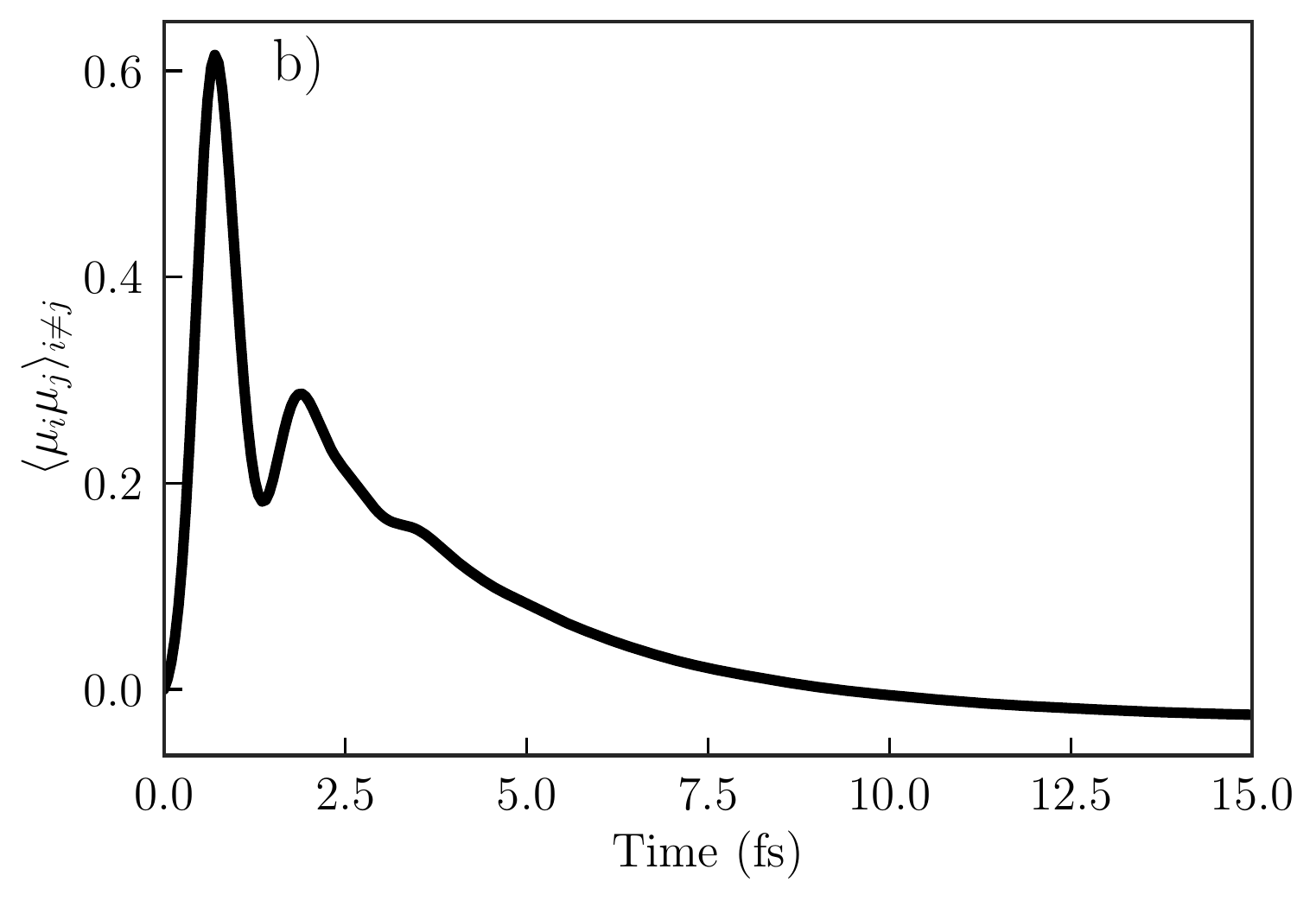}
\caption{\label{fig:4} (a) Expectation value of the
dipole-dipole interaction term, $\langle \mu_{i}\mu_{j} \rangle$ for $i \neq j$
for the simulation shown in \autoref{fig:2}(a) (with $N=5$), within the second
quantized approach (full lines) and without using the permutational symmetry
(dashed lines). (b) The same for the simulation in \autoref{fig:2}(b) with
$N=17$ emitters, only using the second quantized approach.}
\end{figure}

To give another numerical example to illustrate this mapping, we formulate a
simple model Hamiltonian of $N$ $d$-level systems, where the levels of each
emitter are equally separated in energy by $\omega_{\mathrm{e}}$, coupled to a
cavity mode with frequency
$\omega_{\mathrm{c}}=\omega_{\mathrm{e}}=1\,\mathrm{eV}$. The transition
operator of each emitter is defined as $\mu_{i}=\mu_{i}^{+}+\mu_{i}^{-}$, where
$\mu_{i}^{+}=\sum_{\nu=1}^{d-1}\sigma_{\nu,\nu+1}^{i}$ and
$\mu_{i}^{-}=\left(\mu_{i}^{+}\right)^{\dagger}$ and each emitter is coupled to
the cavity mode by a coupling strength of $g=\frac{0.15}{\sqrt{N}}\,\mathrm{eV}$
in the rotating wave approximation. We also include an all-to-all dipole-dipole
interaction term, $H_{\mathrm{d-d}} = D\sum_{k,j\neq k}^{N}\mu_{i}\mu_{j}$,
where $D=0.1\,\mathrm{eV}$. The Hamiltonian is then given by 
\begin{multline}
H = \sum_{i=1}^{N}\sum_{\nu=1}^{d}\omega_{\mathrm{\nu}}\sigma_{\nu,\nu}^{i}+\omega_{\mathrm{c}}a_{\mathrm{c}}^{\dagger}a_{\mathrm{c}}+H_{\mathrm{d-d}}\\
  + g\sum_{i=1}^{N}\left(\mu_{i}^{-}a_{\mathrm{c}}^{\dagger}+\mathrm{H.c.}\right),
\end{multline}
where $\omega_{\mathrm{\nu}} = \nu\omega_{\mathrm{e}}$. The system is under the
action of the incoherent decay of the cavity, $C_{\mathrm{cav}} =
\sqrt{\gamma_{\mathrm{c}}} a_{c}$ where $\gamma_{\mathrm{c}} = 0.15\,$eV, as
well as collective spontaneous emission, $C_{\nu} = \sqrt{\Gamma_{\downarrow}}
\sum_{i=1}^{N}\sigma_{\nu,\nu+1}^{i}$, where $\nu$ runs from $1$ to $d-1$ and
$\Gamma_{\downarrow}=0.05\,\mathrm{eV}$. The initial state is chosen to be the
fully inverted state, $|\psi_{0}\rangle = \prod_{j=1}^{N} |d\rangle_{j} \otimes
|\chi\rangle_{\mathrm{cav}}$, where all emitters are in the most excited state
and the cavity is in the vacuum state. 

The above Hamiltonian and collapse operators are clearly invariant under any
permutation of the emitters. Also, the initial state belongs to the totally
symmetric subspace. Therefore, we can again proceed with the mapping by second
quantizing all relevant operators. In particular, the Hamiltonian can be written
as
\begin{multline}
H = \sum_{\nu=1}^{d}\omega_{\mathrm{\nu}}b_{\nu}^{\dagger}b_{\nu}+\omega_{\mathrm{c}}a_{\mathrm{c}}^{\dagger}a_{\mathrm{c}} 
  + H_{\mathrm{d-d}}\\
  + g\sum_{i=1}^{N}\sum_{\nu=1}^{d}\left(b_{\nu}^{\dagger}b_{\nu+1}a_{\mathrm{c}}^{\dagger}+\mathrm{h.c.}\right),
\end{multline}
where
\begin{multline}
H_{\mathrm{d-d}} = D\sum_{\nu=1}^{d-1}\sum_{\mu=1}^{d-1} \left(b_{\nu}^{\dagger}b_{\mu}^{\dagger}b_{\nu+1}b_{\mu+1}+b_{\nu+1}^{\dagger}b_{\mu+1}^{\dagger}b_{\nu}b_{\mu}\right.\\
  \left.+b_{\nu+1}^{\dagger}b_{\mu}^{\dagger}b_{\nu}b_{\mu+1}+b_{\nu}^{\dagger}b_{\mu+1}^{\dagger}b_{\nu+1}b_{\mu}\right).
\end{multline}
Note that this operator is expressed using normal ordering. The $d-1$ collective
spontaneous emission collapse operators can be rewritten as $C_{\nu} =
\sqrt{\Gamma_{\downarrow}}b_{\nu}^{\dagger}b_{\nu+1}$. Finally, the initial
state is just $|\psi_{0}\rangle = \frac{1}{\sqrt{N!}}
(b_{d}^{\dagger})^{N}|\mathrm{vac}\rangle_{\mathrm{em}}
|\mathrm{vac}\rangle_{\mathrm{cav}}$. 

In \autoref{fig:2} and \autoref{fig:4}, we show the results of the dynamics for
the case of 3-level systems, i.e. $d=3$. The total population of the different
levels, $\left\langle \sum_{i=1}^{N}\sigma_{\nu,\nu}^{i}\right\rangle$, can be
mapped in the second quantized approach to $\left\langle
b_{\nu}^{\dagger}b_{\nu}\right\rangle$. \autoref{fig:2}(a) shows the results for
$N=5$. In \autoref{fig:2}(b), we show the results for $N=17$ emitters. In \autoref{fig:4}, we show the time-dependent expectation value of the dipole-dipole interaction term, $\langle \mu_i \mu_j \rangle$ for $i \neq j$.

Again, as expected, the second quantized approach is completely equivalent to
the direct solution. In this case, the brute-force approach is numerically
intractable, as the number of entries in the density matrix is
$d^{2N}N_{\mathrm{c}}^{2}$, where $N_{\mathrm{c}}$ is the dimension of the
cavity Hilbert space. When working only with the totally symmetric subspace, the
number of entries in the density matrix is reduced to
\begin{equation}
\left(\frac{\left(N+d-1\right)!}{N!\left(d-1\right)!}\right)^{2}N_{\mathrm{c}}^{2},
\end{equation}
greatly reducing the size of  the dynamical object. When comparing with the
approach that uses the symmetrized Liouville space, where the number of entries
in the density matrix is
$\frac{\left(N+d^{2}-1\right)!}{N!\left(d^{2}-1\right)!}N_{\mathrm{c}}^{2}$
\citep{gegg2016efficient,gegg2017identical}, we also get a substantial
reduction. As an example, for $d=3$ and $N=17$, we have a reduction of the
number of entries in the density matrix by a factor of $37$.

Since we start in the fully inverted state, the dipole-dipole interaction,
$H_{\mathrm{d-d}}$, starts to transfer population from the highest excited
emitter state to the intermediate excited state and to a smaller extent to the
emitter ground state. After this first moment, the cavity starts to become
populated and due to its decay, drives the system to its overall ground state,
see \autoref{fig:2}. It is important to notice that due to the dipole-dipole
interaction, $H_{\mathrm{d-d}}$, the ground state of the system is not the state
where all emitters are in their bare ground state. Consequently, there is a
residual population of the intermediate excited state for long times, see
\autoref{fig:2}. This is also the reason why the dipole-dipole interaction goes
to negative values for long times, see \autoref{fig:4}.

\subsection{$N$-excitation subspace}

When working with the dynamics of emitters coupled to cavity modes, there are
situations in which we are not interested in working with the full excitation
subspace. Indeed, in many common cases, restricting to the first or second
excitation subspace is enough~\citep{galego2015cavity, feist2018polaritonic}.
Implementing such a restriction within the current approach is rather simple, as
one only needs to define an operator that determines the number of excitations
in terms of creation and annihilation operators of the emitter levels. As an
example, if one is working with the Holstein-Tavis-Cummings model, where each
emitter is described as having two electronic states, ground and excited, with
one vibrational mode, one could define a subspace where restrictions are imposed
on either the electronic or nuclear excitations, or both.

As a concrete example, we discuss vibrational strong coupling for the case where
a single (approximately harmonic) vibrational mode per molecule is in resonance
with a cavity mode. The simplest Hamiltonian to model collective vibrational
strong coupling is \citep{del2015quantum}
\begin{equation}
H=\omega_{\mathrm{c}}a_{c}^{\dagger}a_{c}+\sum_{i=1}^{N_{\mathrm{mol}}}\omega_{\mathrm{v}}c_{i}^{\dagger}c_{i}+\sum_{i=1}^{N_{\mathrm{mol}}}g\left(a_{c}^{\dagger}c_{i}+\mathrm{H.c.}\right),\label{eq:VSC_HAM1}
\end{equation}
where $a_{\mathrm{c}}$ is the annihilation operator for the cavity mode with
frequency $\omega_{\mathrm{c}}$, and $c_{i}$ is the annihilation operator of the
optically active vibrational mode of molecule $i$, characterized by its
frequency $\omega_{\mathrm{v}}$. $N_{\mathrm{mol}}$ is the number of molecules,
and the cavity-phonon interaction is given by $g$. Rewriting the vibrational
operators using the eigenstates of the harmonic oscillator,
$c_{i}^{\dagger}=\sum_{n=0}^{\infty}\sqrt{n+1}\left|n+1\right\rangle
_{i}\left\langle n\right|_{i}$, \autoref{eq:VSC_HAM1} can be written as
\begin{multline}
H = \omega_{\mathrm{c}}a_{c}^{\dagger}a_{c}+\sum_{i=1}^{N_{\mathrm{mol}}}\sum_{n=0}^{\infty}n\omega_{\mathrm{v}}\left|n\right\rangle _{i}\left\langle n\right|_{i}\\
 + \sum_{i=1}^{N_{\mathrm{mol}}}\sum_{n=0}^{\infty}g\left(a_{c}\sqrt{n+1}\left|n+1\right\rangle _{i}\left\langle n\right|_{i}+\mathrm{h.c.}\right).
\end{multline}
This Hamiltonian is permutationally invariant under the exchange of any two
molecules. If the initial state is in the totally symmetric subspace, we can map
the Hamiltonian to 
\begin{multline}
H = \omega_{\mathrm{c}}a_{c}^{\dagger}a_{c}+\sum_{n=0}^{\infty}n\omega_{\mathrm{v}}b_{n}^{\dagger}b_{n}\nonumber \\
  + \sum_{n=0}^{\infty}g\left(a_{c}\sqrt{n+1}b_{n+1}^{\dagger}b_{n}+\mathrm{h.c.}\right),
\end{multline}
where $b_{n}$ is the bosonic annihilation operator for the state
$\left|n\right\rangle $ and the states of interest are restricted to the
subspace where $\left\langle
\sum_{n=0}^{\infty}b_{n}^{\dagger}b_{n}\right\rangle =N_{\mathrm{mol}}$. For
regimes in which $\omega_{\mathrm{v}}\approx\omega_{\mathrm{c}}$, it is
reasonable to work in the $N_{\mathrm{exc}}$-excitation subspace
\citep{campos2021generalization}. In this formalism, this additional restriction
can be simply formulated as $\left\langle \sum_{n=0}^{\infty} n b_{n}^{\dagger}
b_{n} + a_{c}^{\dagger} a_{c}\right\rangle = N_{\mathrm{exc}}$.

\section{Conclusion\label{sec:Conclusion}}

To conclude, we have proposed a scheme to fully exploit the permutational
symmetry of identical, but arbitrary emitters when only collective dissipation
operators are considered. This scheme relies on the fact that the totally
symmetric subspace is equivalent to a bosonic many-body state. After mapping all
relevant operators to a second quantized picture using a simple procedure, the
explicit construction of the totally symmetric subspace from direct state
products is not required anymore. This approach thus provides a straightforward
and easily implemented way to treat such systems while fully exploiting their
permutational symmetry to significantly reduce the size of the Hilbert space. We
discuss several examples, such as the Tavis-Cummings model, the
Holstein-Tavis-Cummings model and a model Hamiltonian where two-body operators
are taken into account, and explicitly demonstrate the equivalence of the second
quantized approach to direct solution.

We expect that this work will be helpful for simulations that can fully exploit
the permutational symmetry of emitters in totally symmetric cases in a very
simple way. This can be especially useful for situations where each emitter must
be considered as having an internal structure that goes beyond the two-level
approximation, such as necessary in the field of molecular polaritonics. For
such systems, the current approach can provide a significant reduction of the
numerical complexity for very little effort.

\section*{Acknowledgments}

This work has been funded by the European Research Council through grant
ERC-2016-STG-714870 and by the Spanish Ministry for Science, Innovation, and
Universities-AEI through grants RTI2018-099737-B-I00, PCI2018-093145 (through
the QuantERA program of the European Commission), and CEX2018-000805-M (through
the María de Maeztu program for Units of Excellence in R\&D). R.~E.~F.~S\@. also
acknowledges support from the fellowship LCF/BQ/PR21/11840008 from ``La
Caixa'' Foundation (ID 100010434) and from the European Union's Horizon 2020
research and innovation programme under the Marie Sklodowska-Curie grant
agreement No. 847648.

\end{document}